\documentclass[aip,apl,groupedaddress,reprint]{revtex4-1} %

\usepackage{graphicx}
\usepackage{epstopdf}
\usepackage{amssymb,amsmath,amsthm}
\usepackage{float}
\usepackage{tabularx}
\usepackage{array}
\usepackage{wrapfig}
\usepackage{color}
\usepackage{transparent}
\usepackage{cases}
\usepackage{ulem}

\usepackage{xcolor}
\usepackage{blindtext}
\definecolor{landanimal}{rgb}{.545,.1,.1}
\colorlet{ocean}{blue!60!black}

\begin{document}
\normalem
\title{Localized modes on a metasurface through multi-wave interactions}

\author{Martin Lott}
\email[]{martin.lott@univ-grenoble-alpes.fr}
\author{Philippe Roux}
\author{L\'eonard Seydoux}
\author{Benoit Tallon}
\affiliation{Univ. Grenoble Alpes, Univ. Savoie Mont Blanc, CNRS, IRD, IFSTTAR, ISTerre, 38000 Grenoble, France}
\author{Adrien Pelat}
\affiliation{Laboratoire d’Acoustique de l’Université du Mans, UMR CNRS 6613, Avenue Olivier Messiaen, 72085 Le Mans, Cedex 09, France}
\author{Sergey Skipetrov}
\affiliation{University Grenoble Alpes, CNRS, LPMMC, 38000 Grenoble, France}
\author{Andrea Colombi}
\affiliation{Department of Civil, Environmental and Geomatic Engineering, Institute of Structural Engineering, Swiss Federal Institute of Technology, Zurich, Switzerland}
\date{\today}

\begin{abstract}
In this paper, we describe the manifestation of localized states through coherent and incoherent analyses of a diffuse elastic wavefield inside a two-dimensional metamaterial made of a collection of vertical long beams glued to a thin plate. We demonstrate that localized states arise due to multi-wave interactions at the beam–plate attachment when the beams acts as coupled resonators for both compressional and flexural resonances on the metasurface. Due to the low-quality factor compressional resonance of the beams, inside the main bandgap the modal density of the system drops to near a high-quality factor flexural resonance of the beams, and blocks the diffusion process of the wavefield intensity. This experiment physically highlights the tight-binding-like coupling in the localized regime for this two-dimensional metamaterial. 
\end{abstract}

\maketitle

\subsection*{Introduction}
In a strongly disordered scattering medium, it is possible to trap waves into energy loops \citep{anderson1958absence}. At the sample size, the trapping induces spatial renormalization of the diffusion coefficient, which highlights the nonconventional transport property of the medium \citep{hu2008localization, cobus2018transverse, aubry2014recurrent, van1999multiple}. The diffusion process can then be stopped without the use of absorption, on a length ζ known as the localization length. At a length scale that corresponds to a few scatterers inside the medium, the intensity spatially spreads in a deterministic way, due to interference effects. This phenomenon is known as modal localization, and it is the local elementary block for the global trapping effect of the wavefield intensity.

Recent two-dimensional studies have directly imaged microwave localized modes inside a disordered microwave cavity \citep{mortessagne2007direct, laurent2007localized}. In this case, the medium consists of a random distribution of resonant scatterers and has a large frequency bandgap. Around some isolated frequencies, the localized modes are clearly visible and spread according to distances roughly equal to the mean distance between the scatterers. 

In this context, so-called locally resonant metamaterials \citep{pendry2000negative, pendry2006controlling, colombi2017elastic, tallon2017impact, page2011metamaterials} are good candidates for studying the transport phenomena of elastic and nonscalar guided waves. Indeed, the coupling between resonators to a homogeneous substrate degenerates the dispersion relationship and creates bandgaps (i.e., Fano-like resonances), in the same way as for electromagnetic waves. At the bandgap edges, the effective wavenumber is typically very high, and each resonator acts as a strong scatterer, and deviates the waves for lengths shorter than the wavelength. 

In this paper, we revisit the experiment from Rupin et al. \citep{rupin2014experimental}, where a two-dimensional cluster of beams attached to a plate was used to create a metasurface. The beams showed both compressional and flexural resonances that were characterized by a low-quality and a high-quality factor, respectively (i.e., wide-spreading vs. narrow resonant peaks). These resonances strongly modify the $A_0$ flexural Lamb waves of the free plate \citep{rupin2015symmetry, colquitt2017seismic, lottlocally}. In these studies, they showed that large bandgaps are associated with compressional resonances that are contaminated by narrow leakage associated with each flexural resonance. As flexural resonances are coupled to the plate through the bending of the $A_0$ Lamb mode, changing to a thinner plate enhanced the flexural bandgap leakage \citep{roux2017new}. This narrow leakage is a potential candidate for localized modes, because it induces a sharp transition zone inside a frequency band that is dominated by a deficit of wave states (namely, the bandgap). 

In addition to the mesoscopic measurements of the wavefield properties, as for the Ioffe-Regel criterion and the spatial intensity distribution function, we measure here the long-range coupling between the resonators themselves as a proxy for the modal density of the system \citep{hildebrand2010statistical}. The experimental set-up described below allows dense and three-dimensional sampling of the wavefield from the plate itself to the top of the beams, which is key for our analysis. Through this exhaustive dataset, we propose a novel matrix-based estimator of the modal density of such a system. Furthermore, we show direct agreement between the resonator motion and the wavefield properties inside the plate substrate, which casts new light on the strong coupling regime that can arise in locally resonant elastic metamaterials. In this two-dimensional system, we obtain a clear physical meaning of the mesoscopic scattering regime based on the local coupling between coupled beam-like resonators. 

\subsection{Experimental set-up}
\begin{figure}
\includegraphics[scale=0.7]{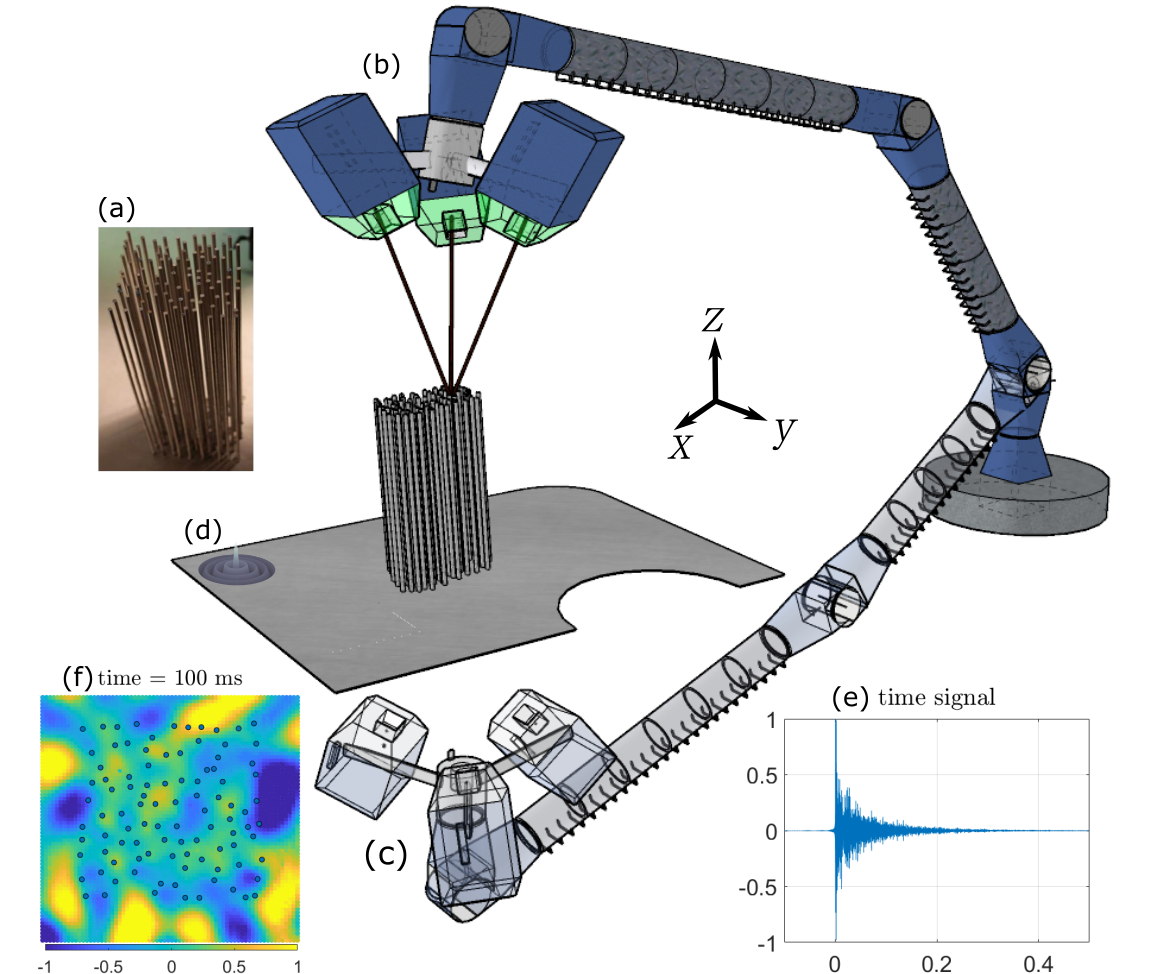}
\caption{\label{fig_1}Scheme of the experiment. The cluster of beams (a) occupies a small area (20 cm × 20 cm) of the 2-mm-thick plate. The three laser heads (b) extract the motions at the top of the beams, as well as on the opposite side of the plate surface through the positioning of the robot arm, depicted with another scanning position in (c). A central computer performs the signal generation and recording, as well as the driving the arm motion. Two piezoelectric sources, one inside and one outside the beam cluster (d), excite the $A_0$ Lamb mode into the plate. The multi-reverberated recorded wavefield (e) includes the plate and beams three-component motion (f).}
\end{figure}

The studied metasurface is a small portion of a thin aluminum plate (2 mm thick, 2 m long, 1 m wide). The experimental scheme is shown in Figure \ref{fig_1}. The plate has a peculiar boundary shape, to break its natural symmetries. A dense set of 98 beams that act as resonators are randomly glued onto the plate surface over an approximate area of 20 cm$^2$, with a mean surface density of 0.5 beam/cm$^2$. Each beam is 60-cm long and has a diameter of 6 mm. The frequency bandwidth considered in the experiment (1-8 kHz) limits the plate waves to the first symmetric and antisymmetric Lamb modes. Due to the plate thickness, most of the energy inside this frequency band propagates as an antisymmetric mode $A_0$ that is characterized by out-of-plane polarization. At around 2.5 kHz, the $A_0$ wavelength is 12 cm, with a wave speed of 300 m/s. The $S_0$ is a lot faster, with a wavelength of 2 m for a wave speed of 5000 m/s. 

The first antisymmetric $A_0$ Lamb wave mode is generated inside the plate using two piezoelectric 1-cm-diameter discs, one of which is located at the center of the beam cluster, and the other in the far-field of the metasurface area. A 3-s-long electrical chirp signal is applied to the piezo source with a frequency ramp starting at 1 kHz, and going up to 8 kHz. The wavefield is then recorded with a three-component laser vibrometer, and cross-correlated with the emitted signal, to provide a 500-ms highly reverberated impulse response with an 80 dB signal-to-noise ratio. 

The measurement platform that is used consists of three scanning laser vibrometers that can measure the three-component motion, which is mounted on a five-axis motorized robot. Through the motion of the robot, we define three scanning areas for which the wavefield is accurately sampled. The first zone concerns the plate in the metamaterial region, on the opposite side of the beam cluster (antisymmetric motion, with regard to the beam cluster side of the plate). The second zone is a free plate region distant from the metamaterial, and the last zone is the top of the beams. The final dataset gathers the exact position in the robot repository and the three components velocity field at each of the scanned points. Due to the expected drop in velocity, the metasurface region is spatially sampled with a 4-mm step size, while the base plate surface is scanned with an 8-mm step size, and the resonator motion with one point per beam. A control unit drives the robot and the laser, and generates and records the signals. 

\subsection{Bandgap and flexural leakage}
\begin{figure}
\includegraphics[scale=0.5]{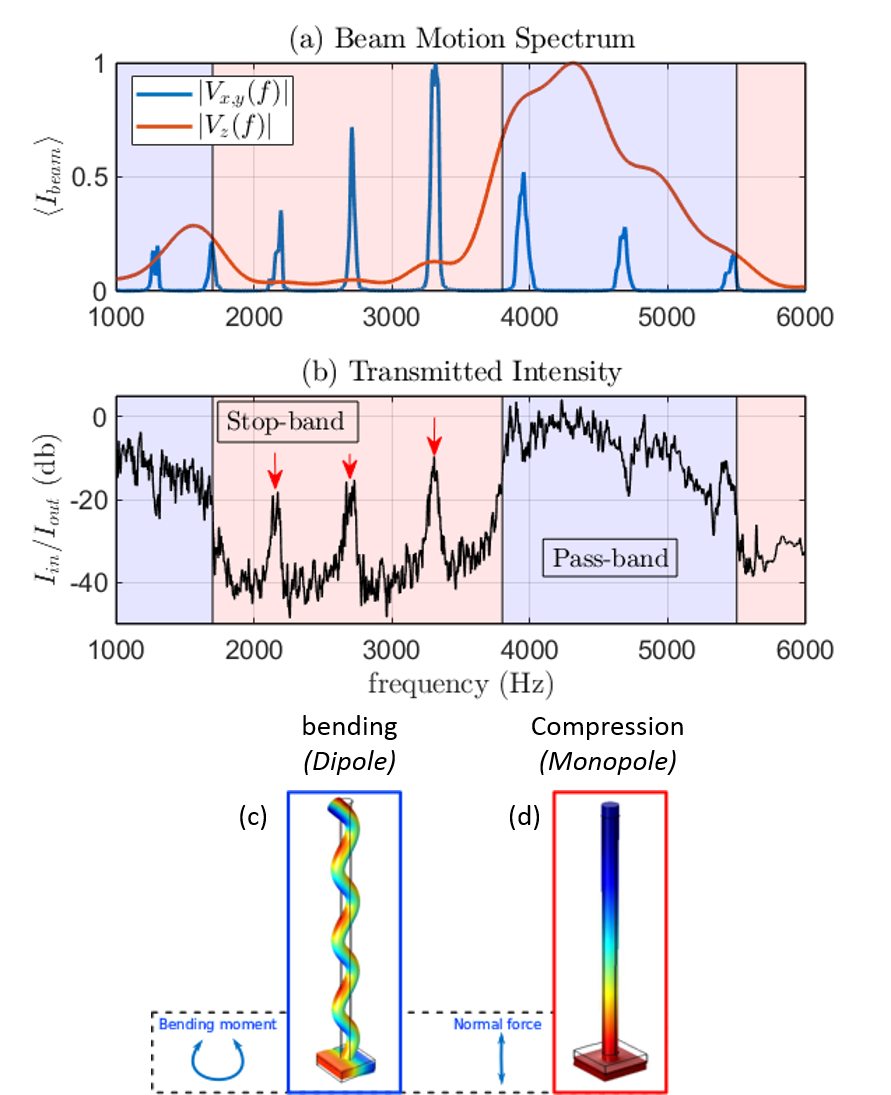}
\caption{\label{fig_2}Metasurface dynamics. (a) Average motion recorded at the top of the beams for compressional (red) and flexural (blue) motions for the source located on the free plate. (b) Transmitted intensity (out-of-plane component) inside the metasurface from the source located outside of the beam cluster. (c, d) Modal representations of the beam motion under flexural (c) and compressional (d) resonance. The resulting stress applied to the plate is either bipolar (bending) or monopolar (compression).}
\end{figure}

figure \ref{fig_2} summarizes the wavefield intensity distribution observed in this experiment. Two different modal vibrations of the beams control the metasurface behavior (fig. \ref{fig_2} c, d). Using the three-component recorded vibration at the top of every beam, the normalized global beam motion is averaged, and the out-of-plan wavefield amplitude $\vert V_z(f) \vert$ is extracted (fig. \ref{fig_2} a, red curve), as well as the in-plane wavefield amplitude $\vert V_{x,y}(f) \vert$ (fig. \ref{fig_2} a, blue curve).

Two dynamics can be clearly noted. On the one hand, the compressional motion with fast pressure waves traveling inside the beam is characterized by a low-quality-factor resonance that spreads over a large frequency band ($\sim$1.7-3.8 kHz). On the other hand, the flexural resonance has slow-traveling shear waves inside the beam, which is hallmark of a high-quality factor. 

The first-order interpretation of the plate-plus-beam coupling has already been described in great detail (see \citep{rupin2014experimental, williams2015theory, colquitt2017seismic, lott2019effective}). Briefly, the low-quality factor compressional motion (fig. \ref{fig_2} a, red curve) of the beams opens the hybridization bandgap, as measured through the intensity transmission of the $A_0$ Lamb mode that is emitted from the source outside the metamaterial region, and is measured inside the beam cluster (fig.\ref{fig_2} b) with the out-of-plane component. Inside the bandgap, the beam cluster clamps the plate and acts as a perfect reflector, which allows only the transmission of evanescent waves with a specific attenuation length. In the latter part of the reverberated signal, after a few tens of milliseconds of propagation, the high-quality factor flexural resonances induce narrowband transmitted intensity peaks (fig. \ref{fig_2} b, red arrows) that are mainly seen in the compressional resonance bandgap. 

We then focus on the bandgap itself, between $\sim$1.7 kHz and 3.6 kHz. At each flexural resonance, waves can propagate inside the beam cluster. However, they do not have the same amplitude dynamics as propagative waves inside the passband. Indeed, as show in figure \ref{fig_2} c, d, when the pass-band dispersion effects are associated to the monopolar stress feedback from the beam to the plate, the bandgap leakage (fig. \ref{fig_2}b, red arrows) at flexural frequencies is due to the dipolar bending moment at the beam–plate attachment. In both cases, out-of-plane polarized $A_0$ Lamb waves are generated, and are thus detected with the out-of-plane component of the laser vibrometer. In the following, we focus on the complex dispersion curve of the out-of-plane $A_0$ waves.

\subsection{Ioffe-Regel criterion of localization}
The dispersion curve is experimentally obtained using a two-point correlation method. Through an averaging process performed over all of the set of receivers and a series of time windows in the coda of the reverberated signals, the two-point correlation function converges toward the effective Green’s function of the propagation medium \citep{lott2019effective}. In the present case, the optimal model for the effective medium is that for the two-dimensional Green’s function for $A_0$ waves in the plate, with the possibility that the effective wave vector will be complex. 

\begin{figure}
\includegraphics[scale=0.6]{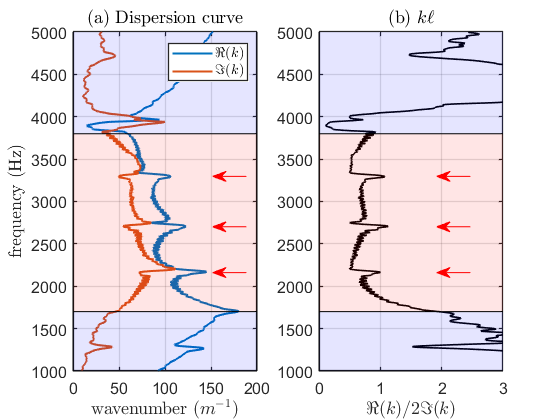}
\caption{\label{fig_3}Complex dispersion curves for the effective medium. (a) Real (blue) and imaginary (red) parts of the wavenumber extracted through the two-point correlation method. (b) Ioffe-Regel “$k\ell$” parameter over frequency, interpreted as the ratio between the real and imaginary parts of the wavenumber. }
\end{figure}

Figure \ref{fig_3}a shows the real (blue) and imaginary (red) parts of the effective wavenumber that were extracted using the two-point correlation method. The real part $\Re(k)$ corresponds to the propagating wavenumber that is connected to the effective wavelength $\lambda=2\pi/\Re(k)$ in the medium. The imaginary part $\Im(k)$ corresponds to an attenuation length $\ell=1/2\Im(k)$ that is classically described as the elastic mean free path in the framework of mesoscopic physics, where multiple scattering has important roles \citep{tallon2017impact}. The first bandgap starts between $\sim$1.5 kHz and 2 kHz and ends just before 4 kHz. On the edge of the pass-band (i.e. $\geq$ 1.5 kHz), the wavelength becomes so small that the beams start to act as strong point-like scatterers. This results in an increase in the imaginary part of the wave vector in the effective Green’s function. Inside the stopband, the real and imaginary parts are of the same order of magnitude, as predicted by Williams et al.\citep{williams2015theory}. The waves are highly attenuated. When the frequency reaches a flexural resonance of the beams, a strong variation in the dispersion curve emerges that reflects the possibility for flexural-type beam excitation to propagate as out-of-plane waves out of the metamaterial area. 

The product between the effective propagating wavenumber and the mean free path represents the Ioffe-Regle criterion of localization  $k\ell\sim 1$, and it describes the relative weight between propagation and damping inside the metamaterial area (fig. \ref{fig_3}b). We observe that $k\ell\gg 1$ in the passband where attenuation is negligible. On the other hand, $k\ell$ strongly decreases in the bandgap, which indicates the strong damping of out-of-plane waves in the metamaterial. We observe that $k\ell$ can reach values of $\leq 1$ within the frequency bandgap, which means that the wave is scattered on a distance less than or equal to one wavelength. In particular, at the flexural resonances,  $k\ell\gg 0.5$  is the first sign of potential localized states in the metamaterial region. 

Like in Laurent \citep{laurent2007localized}, we can evaluate the localization length $\zeta$ from the real and imaginary parts of the wavenumber. We obtain 1 cm at 2205 Hz, 1.3 cm at 2750 Hz, and 1.6 cm at 3350 Hz, which correspond to the lowest distances between the beams. These three distances are also much smaller than the size of the sample (L$\sim$20 cm), which means that localized modes can appear clearly at the frequencies within the random beam configuration. 

\subsection{Intensity distribution}
Localized states are associated with exotic spatial field patterns at the scatterer scale. According to what has been seen before, the waves are scattered at the wavelength scale at the flexural resonances of the beams, and the randomness of the beam distribution starts then to dictate how the intensity of the wavefield spreads spatially. 

Looking at the space and time distribution of the intensity is another way to detect localized states. For transmission through a disordered waveguide, the normalized intensity distribution can be described by Equations (1) and (2) \citep{nieuwenhuizen1995intensity, van1999multiple, hu2008localization}:

\begin{equation}
	\int_{-i\infty}^{i\infty} \dfrac{\mathrm{d}x}{i\pi}K_0(2\sqrt{-ux})\exp(-\phi(x))
\end{equation}
with 
\begin{equation}
	\phi(x) = g\log^2\left(\sqrt{1+\dfrac{x}{g}}+\sqrt{\dfrac{x}{g}}\right)
\end{equation}
where g is the dimensionless conductance that describes the strength of interference effects, and u is the dimensionless intensity ratio, whereby $u=I/\langle I\rangle$. If $g\gg 1$, the waves are diffuse over all of the sample. If $g\sim 1$ the waves can be localized. 

\begin{figure}
\includegraphics[scale=0.55]{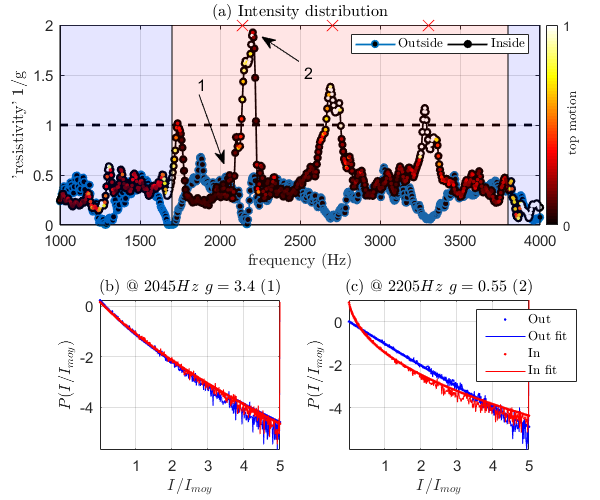}
\caption{\label{fig_4}Intensity distribution: experimental results (b, c) and theoretical fit (a) for the source located on the free plate. (a) The “resistivity” 1/g (g is defined as the Thouless conductance) is plotted with respect to the frequency. The three red crosses indicate the flexural resonance locations, and the color bar indicates the average horizontal motion amplitude of the beams. The two arrows refer to two nearby frequencies with low and high resistivity for which the intensity distribution $I/\langle I \rangle$ is plotted in (b) and (c). (b, c) Blue, curves correspond to Rayleigh-like intensity distribution measured outside of the metamaterial area; red, curves refer to the intensity distribution measured inside the metamaterial. The conductivity g is extracted from a theoretical fit from Equations (1) and (2). (b) is obtained at 2045 Hz, inside the bandgap, and (c) is obtained for a flexural resonance at 2205 Hz.}
\end{figure}

In our set-up, we investigate the distribution of the speckle intensity inside and outside the metamaterial region, and far from the plate boundary. We make the assumption of an ergodic dynamical system. For a 15-ms sliding time window, we compute the ratio $I/\langle I\rangle$ at each frequency, and we gather the intensity ratio over the whole set of time windows in the coda and receiver positions inside the metasurface. The intensity distribution is plotted in figure \ref{fig_4}b, c, both outside and inside the metasurface, for two frequencies that belong to the first bandgap. Deviation in the curve shape from the Rayleigh-like free-plate behavior is clearly seen close to the flexural resonance, at around 2200 Hz. 

The intensity distribution can be fitted to a theory according to the “conductance” g (Eqs. (1), (2)), as shown in figure \ref{fig_4}b, c. Instead of the conductance g, we prefer to plot the “resistivity” of $1/g$ in figure \ref{fig_4}a for the speckle intensity patterns captured both inside and outside the beam cluster. The color bar represents a normalized version of the in-plane average motion at the top of the beams, which is expected to reach a maximum at each flexural resonance. This confirms that the “resistivity” gets stronger when the beams are excited in their transverse motion at flexural resonances inside the bandgap. 

Indeed, the out-of-plane $A_0$-like plate motion inside the bandgap is prevented by the beam anti-resonance compressional motion that “blocks” the propagating plate wave outside the metamaterial. However, this condition is no longer true for local bending motions at each beam–plate attachment. As such, blocking the out-of-plane motion does not prevent a bending moment at the bottom of the beams (fig. \ref{fig_2}b), which becomes overriding at a flexural resonance. We note that this local coupling appears to decrease along the bandgap, with the resistivity parameter $1/g$ slowly decreasing from one flexural motion to the other (e.g., see fig. \ref{fig_4}a, for the three peaks of the resistivity parameter between 1.7 kHz and 3.7 kHz). Indeed, as shown in figure \ref{fig_2}b, the coupling with the flexural motion of the beams is through the bending moment of the plate, which is proportional to the square of the effective wavenumber, and which slowly decreases from 2 kHz to 4 kHz along the bandgap (fig. \ref{fig_3}a). 

Finally, we observe that the flexural resonance inside the passband (fig. \ref{fig_4}a, at $\sim$ 1.3 kHz) does not reveal such behavior. Indeed, whereas $A_0$ waves in the pass-band can still travel inside the metamaterial area, the beams only communicate through tight-binding mechanisms associated with evanescent waves in the bandgap \citep{yves2017crystalline}. 

\subsection{Visualization and robustness of the localized modes }
\begin{figure}
\includegraphics[scale=0.55]{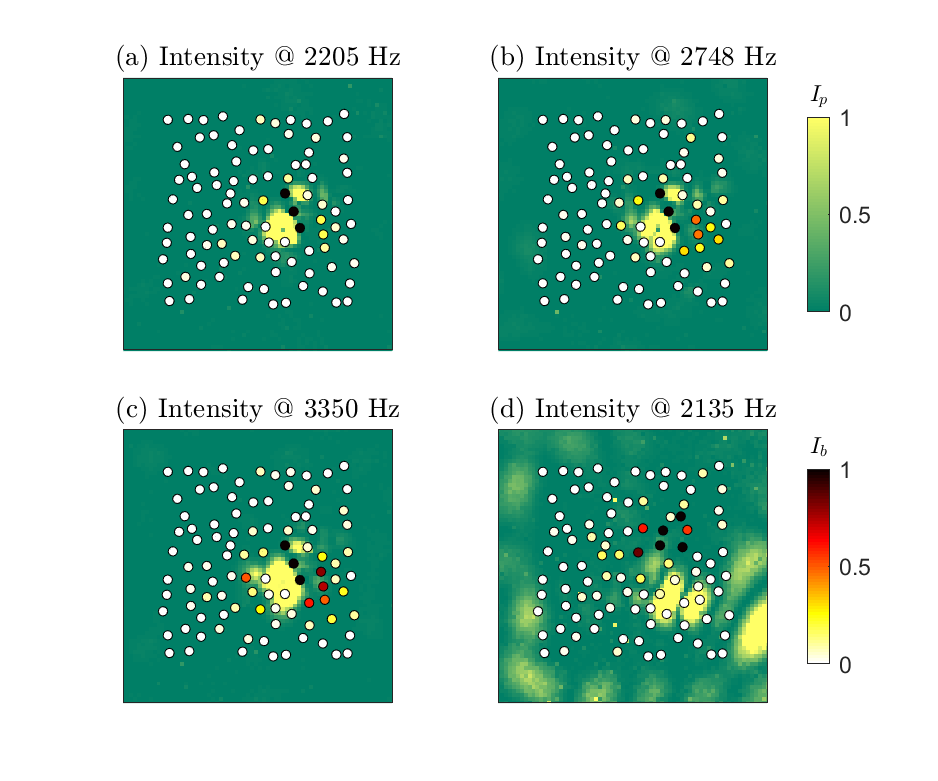}
\caption{\label{fig_5}Normalized intensity maps of the late coda of the signals measured for the source located inside the metamaterial. The horizontal (x-y) beam motion Ib is plotted with the white-to-black color map, and the out-of-plane (z) plate motion Ip is plotted with the green-to-yellow color map. (a-c) The three maxima of the resistivity « 1/g » (shown in Fig. 4a as red crosses, and in Fig. 6a as red arrows). and panel (d) The close frequency of the first maximum where the intensity is leaking outside of the metamaterial area (Fig. 6a, black arrow). }
\end{figure}

Localized states should be robust in terms of the boundary conditions. Here, rather than changing the size of the metamaterial region, as did Mortessagne et al. (2007), we first visually check the robustness of the mode with the source located inside the metamaterial. If the waves are trapped by a localized state, the spatial representation of the intensity in the late part of the reverberating coda should be independent of the recorded time window. Moreover, the energy should be located around the source position (which is located in the middle of the sample), and spread over a small distance (i.e., localization length of $\sim$1.2 cm, as estimated above). Figure \ref{fig_5} shows the intensity inside the metamaterial as one color map that has the beam position and their corresponding flexural intensity motion superimposed as another color map. We used a 0.4-s time window, and started after 0.1 s of propagation for this analysis. Around the three frequencies that correspond to the minimum conductance inside the bandgap (fig. \ref{fig_4}a, red crosses), the intensity is strictly located inside the metamaterial region without any interaction with its boundary, and thus with no leakage into the plate, as shown in figure \ref{fig_5}a-c. For a frequency next to the first flexural resonance inside the bandgap (fig. \ref{fig_5}d), we observe leakage outside the metamaterial. Moreover, as expected through the resistivity 1/g (fig. \ref{fig_4}), the spatial spreading of the three localized modes (fig. \ref{fig_5}a-c) visually increases along the bandgap, in agreement with the effective wavelength (see fig. \ref{fig_3}a).

Another way to quantify the existence of localized modes associated with tight-binding coupling is to investigate the correlation between the beam group motion inside the bandgap. As indicated above, if the beams are distant from each other, they cannot transmit waves one to the other, as the plate is clamped by the collective beam behavior in the bandgap over the metamaterial region. However, when the beams get close to each other, they start to exchange their energy through evanescent waves. One way to evaluate this short-distance correlation is through a cross-spectral matrix, which is computed with the three components of the 98 beams that form the metamaterial. This correlation is defined as given in Equation (3):

\begin{equation}
	C_{ij}(\omega) = q_i(\omega)q_j^\star(\omega)
\end{equation}

where $q_i$ is the three components ($v_x$, $v_y$, $v_z$) of the beams, and $\star$ indicates the complex conjugate. Such a correlations matrix is Hermitian, with $98 \times 3 = 294$ lines and columns. We built the average cross-spectral density matrix (CSDM) as for the two-point correlation method, on an overall average over several realizations of the wavefield, as associated with different time windows of the highly reverberated signals. The selected time windows need to be sufficiently long to include waves that travel over the whole set of beams, from bottom to top, and from top to bottom. As the flexural waves in the beams are much slower than the compressional waves, a 75-ms time window is chosen, which leads to an ensemble average of over 20 realizations. 

\begin{figure}
\includegraphics[scale=0.47]{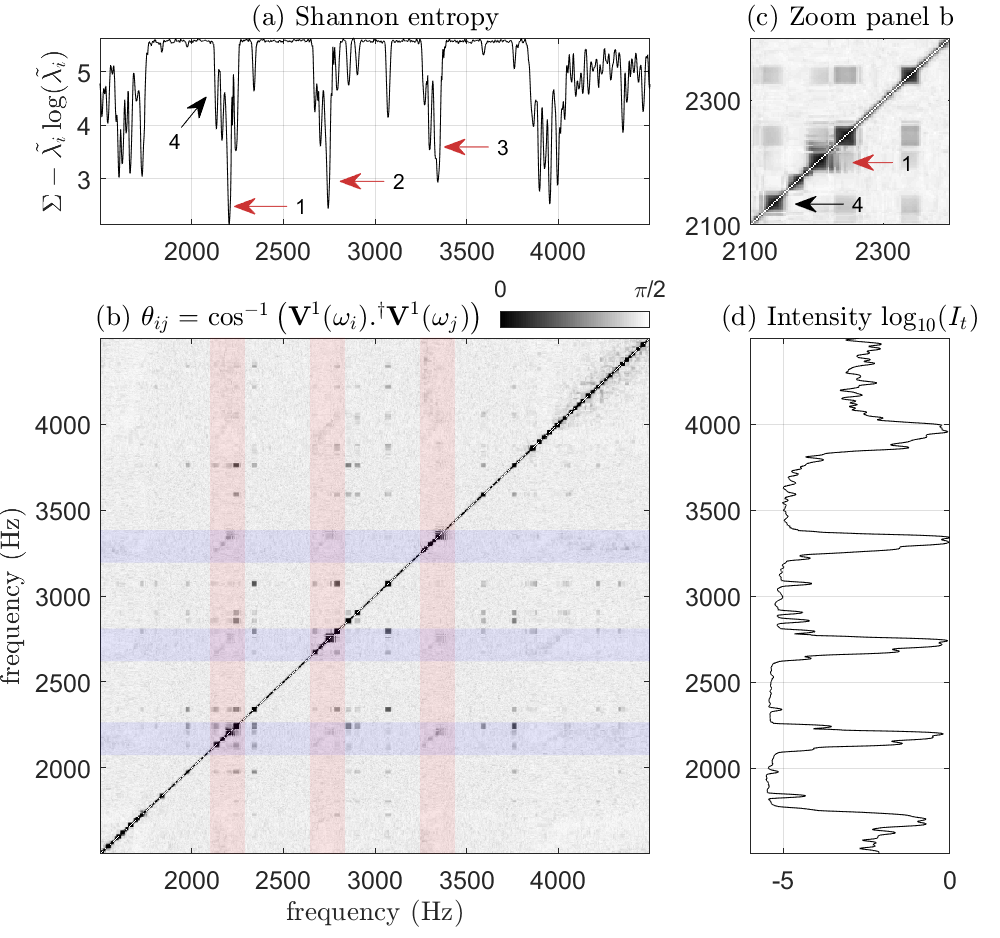}
\caption{\label{fig_6}Singular value decomposition of the cross-spectral density matrix (CSDM) of the three-component motion of the 98 beams. The Shannon entropy (a) is computed from the distribution of the CSDM eigenvalue at each frequency. The three red arrows (1-3) indicate the three minima of the entropy that are associated with the three flexural resonances, which correspond to Figure 5a-c. (b) The scalar product between the first eigenvector extracted between neighboring frequencies shows the robustness of the modal excitation, with magnification around the first flexural resonance (2200Hz) shown in (c). The blue and red shaded areas highlight the flexural resonances of the beams inside the bandgap (see Fig. 2a). (c) The black arrow (4) indicates another mode for the beam cluster, which corresponds to (d) in Figure 5. (d) The average three-component intensity It measured at the top of the beam.}
\end{figure}

If the beam motions are completely independent one from the other, the maximum number of accessible states is reached for the system. If coupling between beams appears, the eigenvalue distribution of the CSDM is affected according to \citep{seydoux2017pre, aubry2014recurrent}. The eigenvalue distribution, which is also known as Shannon entropy, reflects the quantity of information a system can hold; namely, the modal density. This can be quantified at each frequency with the Shannon entropy S defined as in Equation (4):

\begin{equation}
	S = -\Sigma \tilde{\lambda_i}\log(\tilde{\lambda_i})
\end{equation}

where $\tilde{\lambda_i}$ is the i-th trace-normalized eigenvalue of the CSDM \citep{aubry2010singular}. When S drops, the waves lose some degrees of freedom in terms of their propagation. Here, because of the wavefield equipartition, a frequency variation in the eigenvalue distribution is due to a variation in the modal density of the system itself \citep{seydoux2017pre}. Figure \ref{fig_6}a shows the frequency-dependent Shannon entropy for our metamaterial. Once the studied frequency reaches the bandgap (i.e., fig. \ref{fig_6}a, after 1.7 kHz), we only measure the noise at the top of the beams. Indeed, there is no energy on average inside the metamaterial, as depicted in figure \ref{fig_6}d. The beams are static and the residual noise is fully incoherent, which corresponds to the maximum available entropy for the given number of realizations. At the flexural resonances, the curve of figure \ref{fig_6}a can be seen to have gaps, which reach values below those in the passband (i.e., before 1.7 kHz). The modal density is then lower near the flexural resonant frequencies than in the free plate, which confirms the presence of localized modes. Note that the three entropy drops related to the three flexural resonances inside the bandgap (fig. \ref{fig_6}, red arrows 1-3) correspond to the field patterns in figure \ref{fig_5}a-c (respectively). 

To be sure that these drops come with specific and robust modal shapes, we use a correlation criterion \citep{brincker2001modal, gueguen2005soil, michel2010full}. Around each flexural resonant frequency, we look at the modal shape correlation between neighboring frequencies. In the frequency domain, this is equivalent to a scalar product on the eigenvectors that are obtained through the singular value decomposition of the CSDM. In figure \ref{fig_6}a, we represent the “acos” value of the scalar product (which indicates the angle between the two vectors i and j; $\theta_{ij}$) between the first eigenvector $V_1(\omega)$ computed at each frequency from 1.5 kHz to 4.5 kHz. We can see that around each flexural resonance, the off-diagonal terms of the computed matrix confirm the correlation between specific modes at different frequencies, which is clearly highlighted in the magnification of the first flexural resonance from figure \ref{fig_6}b that is shown in figure \ref{fig_6}c. Other groups of modes appear, but not all of them are related to localized states inside the metamaterial. For example, in figure \ref{fig_6}c, the black arrow indicated as 4 refers to the intensity field of figure \ref{fig_5}d, where leakage to the free-plate medium was clearly observed. 

\subsection{Discussion}
Using the computed matrix $\theta_{ij}$, we estimate the modal density of the system as the mode separation $\Delta\omega$ (i.e., distance between elements along the diagonal) and their respective spectral width $\delta\omega$ (i.e., thickness of the diagonal). The Thouless criterion is expressed as $\delta\omega/\Delta\omega$, and correspond to the conductance of the system. In figure \ref{fig_6}c, five “modes” appear between 2100 Hz and 2300 Hz, with a distance between them that is twice that of their respective spectral widths. This means that as expected from the conductance analysis performed from plate measurements (fig. \ref{fig_4}), we find a conductance $g\sim 0.5$ in this frequency band that is based on the resonator modal coupling analysis. 

In theory, there is no mobility edge for two-dimensional systems \citep{cobus2018transverse}, and to get a clear signature of a localized state, we only need a localization length that is a lot shorter than L of the sample. It might be asked why this appears here in particular, at the flexural resonances inside the bandgap, and not somewhere else. Two reasons might be invoked here. First, the modal density of a randomly distributed locally resonant metamaterial shows strong fluctuations near the passband edges, which correspond here to the transmitted bands (due to the flexural resonances) inside the main bandgap \citep{skipetrov2019finite}. Secondly, as specified by \citep{sigalas1996localization}, the effects of randomness are perceived differently according to the wave polarization. In particular, the in-plane motion in the study thereof highlights more sensitivity to randomness (i.e., equivalent to the beam flexural motion here) than to the out-of-plane motion (i.e., equivalent to the beam compressional motion). This makes sense with the present study: on the band edge of each flexural resonance inside the bandgap, the randomness of the metamaterial distribution is strongly enhanced, which produces the appearance of a localized state inside the metamaterial. 

At some point, the mesoscopic response of the wavefield around these localized modes is only driven by the random spatial distribution of the beams attached to the plate. The Ioffe-Regel criterion value of 0.5 for the three localized modes highlighted in this experiment suggests a diffusion regime where a propagating wave is scattered on half of a wavelength. This makes sense with the dipolar diffusion process of one beam around its flexural resonance (fig. \ref{fig_2}c). Inside the bandgap, the beam bases that are attached to the plate cannot move vertically, but can rotate, which is indeed possible every half wavelength.

\subsection*{Conclusion}
In conclusion, we have linked the appearance of localized states to the multi-wave interactions of nearby beams that act as coupled resonators for the plate-plus-beam metasurface. Inside the large main bandgap that is due to the low-quality factor compressional motion of the beams, high-quality factor flexural resonances of the beams make localized modes appear. For frequencies that correspond to the three different flexural resonances inside the main bandgap, the value $k\ell\sim$1 of the effective medium is in agreement with low conductance values (g <1) computed from the wavefield intensity pattern. At these frequencies, the flexural beam motion matches the out-of-plane plate wavefield, which confirms that the neighboring beams transmit energy through evanescent waves. The localization length then appears to be directly related to the effective wavelength inside the main bandgap. We also analyzed the modal density of the system through eigenvalue decomposition of the CSDM computed from the beam motion, which shows that localized modes are robust with respect to frequency. This study highlights the tight-binding coupling prevailing between the beams through the evanescent field that exists in a bandgap with a deficit of wave states. \\


\subsection*{Acknowledgments}
The authors are grateful to Mathieu Sécail and Julien Nicolas from Le Mans Université, and to Pascal Audrain from IRT Jules Verne, for their strong contribution to the experiments using the so-called 3Dvib platform located at the ENSIM School of Engineering in Le Mans.

\subsection*{Bibliography}


\begin{thebibliography}{29}%
\makeatletter
\providecommand \@ifxundefined [1]{%
 \@ifx{#1\undefined}
}%
\providecommand \@ifnum [1]{%
 \ifnum #1\expandafter \@firstoftwo
 \else \expandafter \@secondoftwo
 \fi
}%
\providecommand \@ifx [1]{%
 \ifx #1\expandafter \@firstoftwo
 \else \expandafter \@secondoftwo
 \fi
}%
\providecommand \natexlab [1]{#1}%
\providecommand \enquote  [1]{``#1''}%
\providecommand \bibnamefont  [1]{#1}%
\providecommand \bibfnamefont [1]{#1}%
\providecommand \citenamefont [1]{#1}%
\providecommand \href@noop [0]{\@secondoftwo}%
\providecommand \href [0]{\begingroup \@sanitize@url \@href}%
\providecommand \@href[1]{\@@startlink{#1}\@@href}%
\providecommand \@@href[1]{\endgroup#1\@@endlink}%
\providecommand \@sanitize@url [0]{\catcode `\\12\catcode `\$12\catcode
  `\&12\catcode `\#12\catcode `\^12\catcode `\_12\catcode `\%12\relax}%
\providecommand \@@startlink[1]{}%
\providecommand \@@endlink[0]{}%
\providecommand \url  [0]{\begingroup\@sanitize@url \@url }%
\providecommand \@url [1]{\endgroup\@href {#1}{\urlprefix }}%
\providecommand \urlprefix  [0]{URL }%
\providecommand \Eprint [0]{\href }%
\providecommand \doibase [0]{http://dx.doi.org/}%
\providecommand \selectlanguage [0]{\@gobble}%
\providecommand \bibinfo  [0]{\@secondoftwo}%
\providecommand \bibfield  [0]{\@secondoftwo}%
\providecommand \translation [1]{[#1]}%
\providecommand \BibitemOpen [0]{}%
\providecommand \bibitemStop [0]{}%
\providecommand \bibitemNoStop [0]{.\EOS\space}%
\providecommand \EOS [0]{\spacefactor3000\relax}%
\providecommand \BibitemShut  [1]{\csname bibitem#1\endcsname}%
\let\auto@bib@innerbib\@empty
\bibitem [{\citenamefont {Anderson}(1958)}]{anderson1958absence}%
  \BibitemOpen
  \bibfield  {author} {\bibinfo {author} {\bibfnamefont {P.~W.}\ \bibnamefont
  {Anderson}},\ }\href@noop {} {\bibfield  {journal} {\bibinfo  {journal}
  {Physical review}\ }\textbf {\bibinfo {volume} {109}},\ \bibinfo {pages}
  {1492} (\bibinfo {year} {1958})}\BibitemShut {NoStop}%
\bibitem [{\citenamefont {Hu}\ \emph {et~al.}(2008)\citenamefont {Hu},
  \citenamefont {Strybulevych}, \citenamefont {Page}, \citenamefont
  {Skipetrov},\ and\ \citenamefont {van Tiggelen}}]{hu2008localization}%
  \BibitemOpen
  \bibfield  {author} {\bibinfo {author} {\bibfnamefont {H.}~\bibnamefont
  {Hu}}, \bibinfo {author} {\bibfnamefont {A.}~\bibnamefont {Strybulevych}},
  \bibinfo {author} {\bibfnamefont {J.}~\bibnamefont {Page}}, \bibinfo {author}
  {\bibfnamefont {S.~E.}\ \bibnamefont {Skipetrov}}, \ and\ \bibinfo {author}
  {\bibfnamefont {B.~A.}\ \bibnamefont {van Tiggelen}},\ }\href@noop {}
  {\bibfield  {journal} {\bibinfo  {journal} {Nature Physics}\ }\textbf
  {\bibinfo {volume} {4}},\ \bibinfo {pages} {945} (\bibinfo {year}
  {2008})}\BibitemShut {NoStop}%
\bibitem [{\citenamefont {Cobus}\ \emph {et~al.}(2018)\citenamefont {Cobus},
  \citenamefont {Hildebrand}, \citenamefont {Skipetrov}, \citenamefont {van
  Tiggelen},\ and\ \citenamefont {Page}}]{cobus2018transverse}%
  \BibitemOpen
  \bibfield  {author} {\bibinfo {author} {\bibfnamefont {L.}~\bibnamefont
  {Cobus}}, \bibinfo {author} {\bibfnamefont {W.}~\bibnamefont {Hildebrand}},
  \bibinfo {author} {\bibfnamefont {S.}~\bibnamefont {Skipetrov}}, \bibinfo
  {author} {\bibfnamefont {B.}~\bibnamefont {van Tiggelen}}, \ and\ \bibinfo
  {author} {\bibfnamefont {J.}~\bibnamefont {Page}},\ }\href@noop {} {\bibfield
   {journal} {\bibinfo  {journal} {Physical Review B}\ }\textbf {\bibinfo
  {volume} {98}},\ \bibinfo {pages} {214201} (\bibinfo {year}
  {2018})}\BibitemShut {NoStop}%
\bibitem [{\citenamefont {Aubry}\ \emph {et~al.}(2014)\citenamefont {Aubry},
  \citenamefont {Cobus}, \citenamefont {Skipetrov}, \citenamefont
  {Van~Tiggelen}, \citenamefont {Derode},\ and\ \citenamefont
  {Page}}]{aubry2014recurrent}%
  \BibitemOpen
  \bibfield  {author} {\bibinfo {author} {\bibfnamefont {A.}~\bibnamefont
  {Aubry}}, \bibinfo {author} {\bibfnamefont {L.~A.}\ \bibnamefont {Cobus}},
  \bibinfo {author} {\bibfnamefont {S.~E.}\ \bibnamefont {Skipetrov}}, \bibinfo
  {author} {\bibfnamefont {B.~A.}\ \bibnamefont {Van~Tiggelen}}, \bibinfo
  {author} {\bibfnamefont {A.}~\bibnamefont {Derode}}, \ and\ \bibinfo {author}
  {\bibfnamefont {J.~H.}\ \bibnamefont {Page}},\ }\href@noop {} {\bibfield
  {journal} {\bibinfo  {journal} {Physical review letters}\ }\textbf {\bibinfo
  {volume} {112}},\ \bibinfo {pages} {043903} (\bibinfo {year}
  {2014})}\BibitemShut {NoStop}%
\bibitem [{\citenamefont {van Rossum}\ and\ \citenamefont
  {Nieuwenhuizen}(1999)}]{van1999multiple}%
  \BibitemOpen
  \bibfield  {author} {\bibinfo {author} {\bibfnamefont {M.~v.}\ \bibnamefont
  {van Rossum}}\ and\ \bibinfo {author} {\bibfnamefont {T.~M.}\ \bibnamefont
  {Nieuwenhuizen}},\ }\href@noop {} {\bibfield  {journal} {\bibinfo  {journal}
  {Reviews of Modern Physics}\ }\textbf {\bibinfo {volume} {71}},\ \bibinfo
  {pages} {313} (\bibinfo {year} {1999})}\BibitemShut {NoStop}%
\bibitem [{\citenamefont {Mortessagne}\ \emph {et~al.}(2007)\citenamefont
  {Mortessagne}, \citenamefont {Laurent}, \citenamefont {Legrand},
  \citenamefont {Sebbah},\ and\ \citenamefont
  {Vanneste}}]{mortessagne2007direct}%
  \BibitemOpen
  \bibfield  {author} {\bibinfo {author} {\bibfnamefont {F.}~\bibnamefont
  {Mortessagne}}, \bibinfo {author} {\bibfnamefont {D.}~\bibnamefont
  {Laurent}}, \bibinfo {author} {\bibfnamefont {O.}~\bibnamefont {Legrand}},
  \bibinfo {author} {\bibfnamefont {P.}~\bibnamefont {Sebbah}}, \ and\ \bibinfo
  {author} {\bibfnamefont {C.}~\bibnamefont {Vanneste}},\ }\href@noop {}
  {\bibfield  {journal} {\bibinfo  {journal} {Acta Physica Polonica-Series A
  General Physics}\ }\textbf {\bibinfo {volume} {112}},\ \bibinfo {pages} {665}
  (\bibinfo {year} {2007})}\BibitemShut {NoStop}%
\bibitem [{\citenamefont {Laurent}\ \emph {et~al.}(2007)\citenamefont
  {Laurent}, \citenamefont {Legrand}, \citenamefont {Sebbah}, \citenamefont
  {Vanneste},\ and\ \citenamefont {Mortessagne}}]{laurent2007localized}%
  \BibitemOpen
  \bibfield  {author} {\bibinfo {author} {\bibfnamefont {D.}~\bibnamefont
  {Laurent}}, \bibinfo {author} {\bibfnamefont {O.}~\bibnamefont {Legrand}},
  \bibinfo {author} {\bibfnamefont {P.}~\bibnamefont {Sebbah}}, \bibinfo
  {author} {\bibfnamefont {C.}~\bibnamefont {Vanneste}}, \ and\ \bibinfo
  {author} {\bibfnamefont {F.}~\bibnamefont {Mortessagne}},\ }\href@noop {}
  {\bibfield  {journal} {\bibinfo  {journal} {Physical review letters}\
  }\textbf {\bibinfo {volume} {99}},\ \bibinfo {pages} {253902} (\bibinfo
  {year} {2007})}\BibitemShut {NoStop}%
\bibitem [{\citenamefont {Pendry}(2000)}]{pendry2000negative}%
  \BibitemOpen
  \bibfield  {author} {\bibinfo {author} {\bibfnamefont {J.~B.}\ \bibnamefont
  {Pendry}},\ }\href@noop {} {\bibfield  {journal} {\bibinfo  {journal}
  {Physical review letters}\ }\textbf {\bibinfo {volume} {85}},\ \bibinfo
  {pages} {3966} (\bibinfo {year} {2000})}\BibitemShut {NoStop}%
\bibitem [{\citenamefont {Pendry}\ \emph {et~al.}(2006)\citenamefont {Pendry},
  \citenamefont {Schurig},\ and\ \citenamefont
  {Smith}}]{pendry2006controlling}%
  \BibitemOpen
  \bibfield  {author} {\bibinfo {author} {\bibfnamefont {J.~B.}\ \bibnamefont
  {Pendry}}, \bibinfo {author} {\bibfnamefont {D.}~\bibnamefont {Schurig}}, \
  and\ \bibinfo {author} {\bibfnamefont {D.~R.}\ \bibnamefont {Smith}},\
  }\href@noop {} {\bibfield  {journal} {\bibinfo  {journal} {science}\ }\textbf
  {\bibinfo {volume} {312}},\ \bibinfo {pages} {1780} (\bibinfo {year}
  {2006})}\BibitemShut {NoStop}%
\bibitem [{\citenamefont {Colombi}\ \emph {et~al.}(2017)\citenamefont
  {Colombi}, \citenamefont {Craster}, \citenamefont {Colquitt}, \citenamefont
  {Achaoui}, \citenamefont {Guenneau}, \citenamefont {Roux},\ and\
  \citenamefont {Rupin}}]{colombi2017elastic}%
  \BibitemOpen
  \bibfield  {author} {\bibinfo {author} {\bibfnamefont {A.}~\bibnamefont
  {Colombi}}, \bibinfo {author} {\bibfnamefont {R.~V.}\ \bibnamefont
  {Craster}}, \bibinfo {author} {\bibfnamefont {D.}~\bibnamefont {Colquitt}},
  \bibinfo {author} {\bibfnamefont {Y.}~\bibnamefont {Achaoui}}, \bibinfo
  {author} {\bibfnamefont {S.}~\bibnamefont {Guenneau}}, \bibinfo {author}
  {\bibfnamefont {P.}~\bibnamefont {Roux}}, \ and\ \bibinfo {author}
  {\bibfnamefont {M.}~\bibnamefont {Rupin}},\ }\href@noop {} {\bibfield
  {journal} {\bibinfo  {journal} {Frontiers in Mechanical Engineering}\
  }\textbf {\bibinfo {volume} {3}},\ \bibinfo {pages} {10} (\bibinfo {year}
  {2017})}\BibitemShut {NoStop}%
\bibitem [{\citenamefont {Tallon}\ \emph {et~al.}(2017)\citenamefont {Tallon},
  \citenamefont {Brunet},\ and\ \citenamefont {Page}}]{tallon2017impact}%
  \BibitemOpen
  \bibfield  {author} {\bibinfo {author} {\bibfnamefont {B.}~\bibnamefont
  {Tallon}}, \bibinfo {author} {\bibfnamefont {T.}~\bibnamefont {Brunet}}, \
  and\ \bibinfo {author} {\bibfnamefont {J.~H.}\ \bibnamefont {Page}},\
  }\href@noop {} {\bibfield  {journal} {\bibinfo  {journal} {Physical review
  letters}\ }\textbf {\bibinfo {volume} {119}},\ \bibinfo {pages} {164301}
  (\bibinfo {year} {2017})}\BibitemShut {NoStop}%
\bibitem [{\citenamefont {Page}(2011)}]{page2011metamaterials}%
  \BibitemOpen
  \bibfield  {author} {\bibinfo {author} {\bibfnamefont {J.}~\bibnamefont
  {Page}},\ }\href@noop {} {\bibfield  {journal} {\bibinfo  {journal} {Nature
  materials}\ }\textbf {\bibinfo {volume} {10}},\ \bibinfo {pages} {565}
  (\bibinfo {year} {2011})}\BibitemShut {NoStop}%
\bibitem [{\citenamefont {Rupin}\ \emph {et~al.}(2014)\citenamefont {Rupin},
  \citenamefont {Lemoult}, \citenamefont {Lerosey},\ and\ \citenamefont
  {Roux}}]{rupin2014experimental}%
  \BibitemOpen
  \bibfield  {author} {\bibinfo {author} {\bibfnamefont {M.}~\bibnamefont
  {Rupin}}, \bibinfo {author} {\bibfnamefont {F.}~\bibnamefont {Lemoult}},
  \bibinfo {author} {\bibfnamefont {G.}~\bibnamefont {Lerosey}}, \ and\
  \bibinfo {author} {\bibfnamefont {P.}~\bibnamefont {Roux}},\ }\href@noop {}
  {\bibfield  {journal} {\bibinfo  {journal} {Physical review letters}\
  }\textbf {\bibinfo {volume} {112}},\ \bibinfo {pages} {234301} (\bibinfo
  {year} {2014})}\BibitemShut {NoStop}%
\bibitem [{\citenamefont {Rupin}\ \emph {et~al.}(2015)\citenamefont {Rupin},
  \citenamefont {Roux}, \citenamefont {Lerosey},\ and\ \citenamefont
  {Lemoult}}]{rupin2015symmetry}%
  \BibitemOpen
  \bibfield  {author} {\bibinfo {author} {\bibfnamefont {M.}~\bibnamefont
  {Rupin}}, \bibinfo {author} {\bibfnamefont {P.}~\bibnamefont {Roux}},
  \bibinfo {author} {\bibfnamefont {G.}~\bibnamefont {Lerosey}}, \ and\
  \bibinfo {author} {\bibfnamefont {F.}~\bibnamefont {Lemoult}},\ }\href@noop
  {} {\bibfield  {journal} {\bibinfo  {journal} {Scientific reports}\ }\textbf
  {\bibinfo {volume} {5}},\ \bibinfo {pages} {13714} (\bibinfo {year}
  {2015})}\BibitemShut {NoStop}%
\bibitem [{\citenamefont {Colquitt}\ \emph {et~al.}(2017)\citenamefont
  {Colquitt}, \citenamefont {Colombi}, \citenamefont {Craster}, \citenamefont
  {Roux},\ and\ \citenamefont {Guenneau}}]{colquitt2017seismic}%
  \BibitemOpen
  \bibfield  {author} {\bibinfo {author} {\bibfnamefont {D.}~\bibnamefont
  {Colquitt}}, \bibinfo {author} {\bibfnamefont {A.}~\bibnamefont {Colombi}},
  \bibinfo {author} {\bibfnamefont {R.}~\bibnamefont {Craster}}, \bibinfo
  {author} {\bibfnamefont {P.}~\bibnamefont {Roux}}, \ and\ \bibinfo {author}
  {\bibfnamefont {S.}~\bibnamefont {Guenneau}},\ }\href@noop {} {\bibfield
  {journal} {\bibinfo  {journal} {Journal of the Mechanics and Physics of
  Solids}\ }\textbf {\bibinfo {volume} {99}},\ \bibinfo {pages} {379} (\bibinfo
  {year} {2017})}\BibitemShut {NoStop}%
\bibitem [{\citenamefont {Lott}\ and\ \citenamefont {Roux}()}]{lottlocally}%
  \BibitemOpen
  \bibfield  {author} {\bibinfo {author} {\bibfnamefont {M.}~\bibnamefont
  {Lott}}\ and\ \bibinfo {author} {\bibfnamefont {P.}~\bibnamefont {Roux}},\
  }\href@noop {} {\bibinfo  {journal} {Fundamentals and Applications of
  Acoustic Metamaterials}\ ,\ \bibinfo {pages} {25}}\BibitemShut {NoStop}%
\bibitem [{\citenamefont {Roux}\ \emph {et~al.}(2017)\citenamefont {Roux},
  \citenamefont {Rupin}, \citenamefont {Lemoult}, \citenamefont {Lerosey},
  \citenamefont {Colombi}, \citenamefont {Craster}, \citenamefont
  {Gu{\'e}nneau}, \citenamefont {Kuperman},\ and\ \citenamefont
  {Willams}}]{roux2017new}%
  \BibitemOpen
\bibfield  {journal} {  }\bibfield  {author} {\bibinfo {author} {\bibfnamefont
  {P.}~\bibnamefont {Roux}}, \bibinfo {author} {\bibfnamefont {M.}~\bibnamefont
  {Rupin}}, \bibinfo {author} {\bibfnamefont {F.}~\bibnamefont {Lemoult}},
  \bibinfo {author} {\bibfnamefont {G.}~\bibnamefont {Lerosey}}, \bibinfo
  {author} {\bibfnamefont {A.}~\bibnamefont {Colombi}}, \bibinfo {author}
  {\bibfnamefont {R.}~\bibnamefont {Craster}}, \bibinfo {author} {\bibfnamefont
  {S.}~\bibnamefont {Gu{\'e}nneau}}, \bibinfo {author} {\bibfnamefont {W.~A.}\
  \bibnamefont {Kuperman}}, \ and\ \bibinfo {author} {\bibfnamefont {E.~G.}\
  \bibnamefont {Willams}},\ }\href@noop {} {\enquote {\bibinfo {title} {New
  trends toward locally-resonant metamaterials at the mesoscopic scale},}\ }
  (\bibinfo {year} {2017})\BibitemShut {NoStop}%
\bibitem [{\citenamefont {Hildebrand}\ \emph {et~al.}(2010)\citenamefont
  {Hildebrand}, \citenamefont {Cobus},\ and\ \citenamefont
  {Page}}]{hildebrand2010statistical}%
  \BibitemOpen
  \bibfield  {author} {\bibinfo {author} {\bibfnamefont {W.}~\bibnamefont
  {Hildebrand}}, \bibinfo {author} {\bibfnamefont {L.}~\bibnamefont {Cobus}}, \
  and\ \bibinfo {author} {\bibfnamefont {J.}~\bibnamefont {Page}},\ }\href@noop
  {} {\bibfield  {journal} {\bibinfo  {journal} {The Journal of the Acoustical
  Society of America}\ }\textbf {\bibinfo {volume} {127}},\ \bibinfo {pages}
  {2819} (\bibinfo {year} {2010})}\BibitemShut {NoStop}%
\bibitem [{\citenamefont {Williams}\ \emph {et~al.}(2015)\citenamefont
  {Williams}, \citenamefont {Roux}, \citenamefont {Rupin},\ and\ \citenamefont
  {Kuperman}}]{williams2015theory}%
  \BibitemOpen
  \bibfield  {author} {\bibinfo {author} {\bibfnamefont {E.~G.}\ \bibnamefont
  {Williams}}, \bibinfo {author} {\bibfnamefont {P.}~\bibnamefont {Roux}},
  \bibinfo {author} {\bibfnamefont {M.}~\bibnamefont {Rupin}}, \ and\ \bibinfo
  {author} {\bibfnamefont {W.}~\bibnamefont {Kuperman}},\ }\href@noop {}
  {\bibfield  {journal} {\bibinfo  {journal} {Physical Review B}\ }\textbf
  {\bibinfo {volume} {91}},\ \bibinfo {pages} {104307} (\bibinfo {year}
  {2015})}\BibitemShut {NoStop}%
\bibitem [{\citenamefont {Lott}\ and\ \citenamefont
  {Roux}(2019)}]{lott2019effective}%
  \BibitemOpen
  \bibfield  {author} {\bibinfo {author} {\bibfnamefont {M.}~\bibnamefont
  {Lott}}\ and\ \bibinfo {author} {\bibfnamefont {P.}~\bibnamefont {Roux}},\
  }\href@noop {} {\bibfield  {journal} {\bibinfo  {journal} {Physical Review
  Materials}\ }\textbf {\bibinfo {volume} {3}},\ \bibinfo {pages} {065202}
  (\bibinfo {year} {2019})}\BibitemShut {NoStop}%
\bibitem [{\citenamefont {Nieuwenhuizen}\ and\ \citenamefont
  {Van~Rossum}(1995)}]{nieuwenhuizen1995intensity}%
  \BibitemOpen
  \bibfield  {author} {\bibinfo {author} {\bibfnamefont {T.~M.}\ \bibnamefont
  {Nieuwenhuizen}}\ and\ \bibinfo {author} {\bibfnamefont {M.}~\bibnamefont
  {Van~Rossum}},\ }\href@noop {} {\bibfield  {journal} {\bibinfo  {journal}
  {Physical review letters}\ }\textbf {\bibinfo {volume} {74}},\ \bibinfo
  {pages} {2674} (\bibinfo {year} {1995})}\BibitemShut {NoStop}%
\bibitem [{\citenamefont {Yves}\ \emph {et~al.}(2017)\citenamefont {Yves},
  \citenamefont {Lemoult}, \citenamefont {Fink},\ and\ \citenamefont
  {Lerosey}}]{yves2017crystalline}%
  \BibitemOpen
  \bibfield  {author} {\bibinfo {author} {\bibfnamefont {S.}~\bibnamefont
  {Yves}}, \bibinfo {author} {\bibfnamefont {F.}~\bibnamefont {Lemoult}},
  \bibinfo {author} {\bibfnamefont {M.}~\bibnamefont {Fink}}, \ and\ \bibinfo
  {author} {\bibfnamefont {G.}~\bibnamefont {Lerosey}},\ }\href@noop {}
  {\bibfield  {journal} {\bibinfo  {journal} {Scientific reports}\ }\textbf
  {\bibinfo {volume} {7}},\ \bibinfo {pages} {15359} (\bibinfo {year}
  {2017})}\BibitemShut {NoStop}%
\bibitem [{\citenamefont {Seydoux}\ \emph {et~al.}(2017)\citenamefont
  {Seydoux}, \citenamefont {de~Rosny},\ and\ \citenamefont
  {Shapiro}}]{seydoux2017pre}%
  \BibitemOpen
  \bibfield  {author} {\bibinfo {author} {\bibfnamefont {L.}~\bibnamefont
  {Seydoux}}, \bibinfo {author} {\bibfnamefont {J.}~\bibnamefont {de~Rosny}}, \
  and\ \bibinfo {author} {\bibfnamefont {N.~M.}\ \bibnamefont {Shapiro}},\
  }\href@noop {} {\bibfield  {journal} {\bibinfo  {journal} {Geophysical
  Journal International}\ }\textbf {\bibinfo {volume} {210}},\ \bibinfo {pages}
  {1432} (\bibinfo {year} {2017})}\BibitemShut {NoStop}%
\bibitem [{\citenamefont {Aubry}\ and\ \citenamefont
  {Derode}(2010)}]{aubry2010singular}%
  \BibitemOpen
  \bibfield  {author} {\bibinfo {author} {\bibfnamefont {A.}~\bibnamefont
  {Aubry}}\ and\ \bibinfo {author} {\bibfnamefont {A.}~\bibnamefont {Derode}},\
  }\href@noop {} {\bibfield  {journal} {\bibinfo  {journal} {Waves in Random
  and Complex Media}\ }\textbf {\bibinfo {volume} {20}},\ \bibinfo {pages}
  {333} (\bibinfo {year} {2010})}\BibitemShut {NoStop}%
\bibitem [{\citenamefont {Brincker}\ \emph {et~al.}(2001)\citenamefont
  {Brincker}, \citenamefont {Zhang},\ and\ \citenamefont
  {Andersen}}]{brincker2001modal}%
  \BibitemOpen
  \bibfield  {author} {\bibinfo {author} {\bibfnamefont {R.}~\bibnamefont
  {Brincker}}, \bibinfo {author} {\bibfnamefont {L.}~\bibnamefont {Zhang}}, \
  and\ \bibinfo {author} {\bibfnamefont {P.}~\bibnamefont {Andersen}},\
  }\href@noop {} {\bibfield  {journal} {\bibinfo  {journal} {Smart materials
  and structures}\ }\textbf {\bibinfo {volume} {10}},\ \bibinfo {pages} {441}
  (\bibinfo {year} {2001})}\BibitemShut {NoStop}%
\bibitem [{\citenamefont {Gueguen}\ and\ \citenamefont
  {Bard}(2005)}]{gueguen2005soil}%
  \BibitemOpen
  \bibfield  {author} {\bibinfo {author} {\bibfnamefont {P.}~\bibnamefont
  {Gueguen}}\ and\ \bibinfo {author} {\bibfnamefont {P.-Y.}\ \bibnamefont
  {Bard}},\ }\href@noop {} {\bibfield  {journal} {\bibinfo  {journal} {Journal
  of Earthquake Engineering}\ }\textbf {\bibinfo {volume} {9}},\ \bibinfo
  {pages} {657} (\bibinfo {year} {2005})}\BibitemShut {NoStop}%
\bibitem [{\citenamefont {Michel}\ \emph {et~al.}(2010)\citenamefont {Michel},
  \citenamefont {Gu{\'e}guen}, \citenamefont {El~Arem}, \citenamefont
  {Mazars},\ and\ \citenamefont {Kotronis}}]{michel2010full}%
  \BibitemOpen
  \bibfield  {author} {\bibinfo {author} {\bibfnamefont {C.}~\bibnamefont
  {Michel}}, \bibinfo {author} {\bibfnamefont {P.}~\bibnamefont {Gu{\'e}guen}},
  \bibinfo {author} {\bibfnamefont {S.}~\bibnamefont {El~Arem}}, \bibinfo
  {author} {\bibfnamefont {J.}~\bibnamefont {Mazars}}, \ and\ \bibinfo {author}
  {\bibfnamefont {P.}~\bibnamefont {Kotronis}},\ }\href@noop {} {\bibfield
  {journal} {\bibinfo  {journal} {Earthquake Engineering \& Structural
  Dynamics}\ }\textbf {\bibinfo {volume} {39}},\ \bibinfo {pages} {419}
  (\bibinfo {year} {2010})}\BibitemShut {NoStop}%
\bibitem [{\citenamefont {Skipetrov}(2019)}]{skipetrov2019finite}%
  \BibitemOpen
  \bibfield  {author} {\bibinfo {author} {\bibfnamefont {S.~E.}\ \bibnamefont
  {Skipetrov}},\ }\href@noop {} {\bibfield  {journal} {\bibinfo  {journal}
  {arXiv preprint arXiv:1909.13661}\ } (\bibinfo {year} {2019})}\BibitemShut
  {NoStop}%
\bibitem [{\citenamefont {Sigalas}\ \emph {et~al.}(1996)\citenamefont
  {Sigalas}, \citenamefont {Soukoulis}, \citenamefont {Chan},\ and\
  \citenamefont {Turner}}]{sigalas1996localization}%
  \BibitemOpen
  \bibfield  {author} {\bibinfo {author} {\bibfnamefont {M.}~\bibnamefont
  {Sigalas}}, \bibinfo {author} {\bibfnamefont {C.}~\bibnamefont {Soukoulis}},
  \bibinfo {author} {\bibfnamefont {C.-T.}\ \bibnamefont {Chan}}, \ and\
  \bibinfo {author} {\bibfnamefont {D.}~\bibnamefont {Turner}},\ }\href@noop {}
  {\bibfield  {journal} {\bibinfo  {journal} {Physical Review B}\ }\textbf
  {\bibinfo {volume} {53}},\ \bibinfo {pages} {8340} (\bibinfo {year}
  {1996})}\BibitemShut {NoStop}%
\end{thebibliography}
\end{document}